Seamless maps of major elements of the Moon: Results from high-resolution geostationary satellite

Yu Lu[1,2,3], Yun-Zhao Wu[1,4*], Cui Li[1], Jin-Song Ma[5,6], Wen-Wen Qi[6], Wei Tan[7], Xiao-Man Li[7], Zhi-Cheng Shi[7], Hong-Yan He[7], Shu-Wu Dai[8], Guo Li[8], Feng-Jing Liu[8], Jing-Qiao Wang[9], Xiao-Yan Wang[9], Qi Wang[10] and Ling-Jie Meng[10]

[1] Key Laboratory of Planetary Sciences, Purple Mountain Observatory, Chinese Academy of Sciences, Nanjing 210034, China; *wu@pmo.ac.cn*

[2] School of Astronomy and Space Science, University of Science and Technology of China, Hefei 230026, China

[3] CAS Center for Excellence in Comparative Planetology, Hefei 230026, China

[4] Space Science Institute, Macau University of Science and Technology, Macau, China

[5] School of Geography and Ocean Science, Nanjing University, Nanjing 210023, China

[6] Jiangsu Center for Collaborative Innovation in Geographical Information Resource Development and Application, Nanjing 210023, China

[7] Beijing Institute of Space Mechanics and Electricity, Beijing, China

[8] Beijing Institute of Spacecraft System Engineering, Beijing, China

[9] China Centre for Resources Satellite Data and Application, Beijing, China

[10] Earth Observation System and Data Center, China National Space Administration, Beijing, China

* Corresponding author.

Abstract:

Major elements such as Fe, Ti, Mg, Al, Ca, and Si play very important roles in understanding the origin and evolution of the Moon. Previous maps of these major elements derived from orbital data are based on mosaic images or low-resolution Gamma ray data. The hue variations and gaps among orbital boundaries in the mosaic images are not conducive to geological studies. This paper aims to produce seamless and homogenous distribution maps of major elements using the single-exposure image of the whole lunar disk obtained by China's high-resolution geostationary satellite, Gaofen-4, with a spatial resolution of ~500 m. The elemental contents of soil samples returned by Apollo and Luna missions were used as ground truth, and were correlated with the reflectance of the sampling sites extracted from Gaofen-4 data. The final distribution maps of these major oxides are generated with the statistical regression model. With these products the average contents and proportions of the major elements for maria and highlands were estimated and compared. The results showed that $SiO_2$ and $TiO_2$ have the highest and lowest fractions in mare and highland areas, respectively. Besides, the relative concentrations of these elements could serve as indicators of geologic processes, e.g., the obviously asymmetric distributions of $Al_2O_3$, $CaO$, and $SiO_2$ around Tycho crater may suggest that Tycho crater was formed by an oblique impact from the southwest direction.

1. Introduction

Information on the global chemistry of major elements such as Fe, Ti, Mg, Al, Ca, Si are fundamentally significant for understanding the composition, origin, and evolution of the Moon, and thus are important scientific objective in lunar exploration. The classic Lunar Magma Ocean theory believed that the Moon was initially molten. During the cooling of the Moon, dense magnesium silicates olivine and pyroxene crystallized first and sank to the bottom of the magma ocean, forming the lunar mantle. The remaining magma became increasingly rich in calcium and aluminum until plagioclase began to crystallize and float to the surface of the magma ocean, forming the highlands crust (Wood et al. 1970; Warren 1985, 1990; Carlson 2019). Of the six major elements, Fe and Mg are rich in olivine and pyroxene in the mantle, while Al and Ca are abundant in plagioclase in the lunar crust. Mapping the crustal distributions of these major elements is beneficial for understanding the geochemical composition and geological evolution of the bulk Moon.

Lunar samples provide the most direct and accurate information about the elements of the Moon. However, these samples are very limited considering the large scope of the Moon. Remote sensing technology is widely used to explore the spatial distribution of the elements on the Moon. Among these six elements, Fe and Ti are transition elements, which exhibit absorption features and could be quantitatively estimated with optical spectroscopy. The distribution of FeO and $TiO_2$ contents have been derived by many researchers using different data, such as Clementine ultraviolet-visible (UV/VIS) images (Lucey et al. 1995, 2000; Blewett et al. 1997; Gillis et al. 2004, 2006), Hubble Space Telescope data (Robinson et al. 2007), Chang'E-1 Interference Imaging Spectrometer (IIM) data (Wu 2012; Wu et al. 2012; Xia et al. 2019), Chang'E-2 microwave sounder data (Liu et al. 2019), Kaguya Multiband Imager data (Otake et al. 2012; Lemelin et al. 2016), Chandrayaan-1 Moon Mineralogy Mapper ($M^3$) data (Bhatt et al. 2019), and Lunar Reconnaissance Orbiter Camera Wide Angle Camera data (Sato et al. 2017). Although nonchromophore elements, such as Mg, Al, Ca, and Si, don't have diagnostic absorption features, they can also affect the Moon's reflectance values, and thus could be estimated by optical spectroscopy (Fischer & Pieters 1995; Shkuratov et al. 2003, 2005; Wu 2012). Fischer & Pieters (1995) mapped the Al concentrations on the Moon by Galileo solid state imaging system based on the positive correlation between lunar surface reflectance and aluminum concentration. Bhatt et al. (2019) mapped the abundances of Fe, Ca, and Mg using $M^3$ data. Various methods, such as principal component analysis (Jaumann 1991), principal component regression (Pieters et al. 2002), partial least squares regression (Li 2006; Wu 2012), support vector machine (Zhang et al. 2009; Bhatt et al. 2019), second-order polynomial regression model (Wöhler et al. 2011), multivariate linear regression (Bhatt et al. 2019), and neural networks model (Xia et al. 2019), have been applied to estimate these major elements, and the predictions are quite successful.

Optical spectroscopy can be used for producing high resolution maps of major elements. Shkuratov et al. (2005) generated global maps of major elements with Clementine UV/VIS images. Wu (2012) and Xia et al. (2019) produced the maps of the abundances of the six major elements and Mg# (the molar or atom ratio of Mg/(Mg+Fe)) using Chang'E-1 IIM data. However, these maps have mosaic borders due to calibration and photometric artifacts or gaps due to lack of data (e.g., Clementine, $M^3$, and IIM results). The hue variations and gaps among orbital boundaries in the mosaic images are not conducive to geological studies, e.g., separation of different geologic units. This study aimed to produce seamless and homogeneous maps of major elements using the single-exposure image of the whole lunar disk obtained by the high-resolution geostationary satellite, Gaofen-4.

2. Data and method

2.1 Data

Gaofen-4 (GF-4) is the first high-resolution geostationary satellite in China. It has 6 spectral channels: 450~900 nm (band 1) in panchromatic, 450~520 nm (band 2) in blue, 520~600 nm (band 3) in green, 630~690 nm (band 4) in red, 760~900 nm (band 5) in near-infrared, and 3.5~4.1 μm (band 6) in mid-infrared. GF-4 has a large-array VNIR detector with fields of view (FOV) of 0.8°×0.8° and instantaneous fields of view (IFOV) of 1.363 μrad/pixel. At 04:49:00 UTC on July 28, 2018, GF-4 imaged the Moon in visible and near-infrared (bands 1-5) with spatial resolution of ~500 m (Wu et al. 2020) (Fig. 1). (The GF-4 images of Copernicus crater derived at 4 different local times are shown in Fig. S1 as an example). In this observation, the Moon-Sun distance is 1.015 AU, the Moon-Camera distance is 44.581×10$^4$ km, the Sub-solar point is 0.02°S, 4.77°W, the Sub-camera point is 1.47°S, 1.17°W, and the phase angle is 3.88°. This study used band 5 (near-infrared) data to estimate the contents of the major elements since in this band the Moon has relatively higher energy. The lunar effective wavelength for band 5 is 0.81 μm.

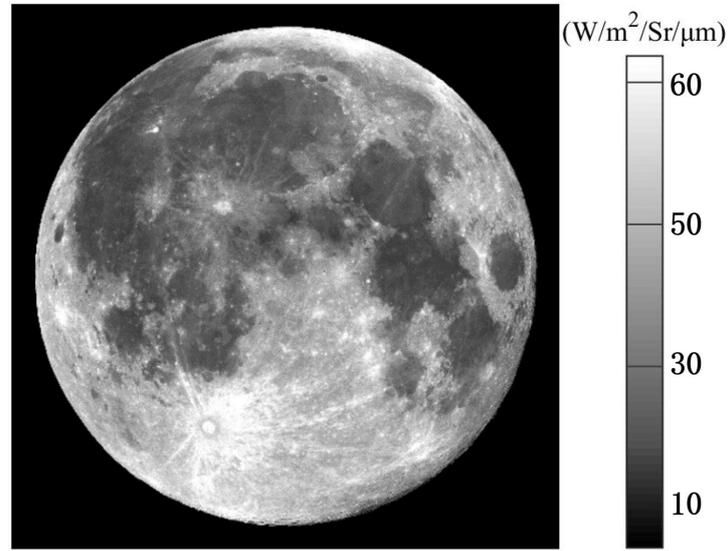

Fig. 1. The radiance of the Moon imaged by the GF-4 band 5 on July 28, 2018 (Wu et al. 2020).

The reflectance data were calculated using the equation below:

$$Ref = \frac{RAD \times \pi \times d^2}{E} \times \frac{\mu_0 + \mu}{\mu_0} \quad (1)$$

where $Ref$ is reflectance, $RAD$ is the radiance, $\mu_0 = \cos i$, $i$ is the solar zenith angle, $\mu = \cos e$, $e$ is the satellite zenith angle, and $d$ represents the distance between the Sun and the Moon. $E$ is the solar irradiance at 1 AU, and was resampled according to the wavelength of GF-4 band 5:

$$E = \frac{\int_{\lambda_1}^{\lambda_2} J_\lambda S_b(\lambda) d\lambda}{\int_{\lambda_1}^{\lambda_2} S_b(\lambda) d\lambda} \quad (2)$$

where $E$ is the resampled solar irradiance of GF-4 band 5, $J_\lambda$ is the original solar flux, and $S_b(\lambda)$ is the spectral response function of the band 5.

The elemental contents of soil samples returned by Apollo and Luna missions were used as ground truth and correlated with the reflectance of the sampling sites extracted from the GF-4 data. In addition to the sample sites used in Wu (2012), the Chang'E-3 site (Wu et al. 2018) and two Apollo 15 sites were also used in this study (Table 1). The sampling station coordinates come from Wu (2012), and detailed elements contents at these sample sites are from the references listed in Table 1. Besides, some of the Apollo 15, 16, and 17 sampling sites can't be resolved in the GF-4 image, thus the elemental contents at some sampling sites were averaged. The reflectance value of each sample station was manually extracted from the GF-4 data considering its latitude, longitude and the published traverse maps.

Table 1. Major elements abundances for lunar sample sites used in this study.

| Site | Lat | Lon | Ref | FeO | TiO$_2$ | MgO | CaO | Al$_2$O$_3$ | SiO$_2$ | sample | references |
|---|---|---|---|---|---|---|---|---|---|---|---|
| Chang'E-3 | 44.13 | 340.49 | 0.18 | 22.24 | 4.31 | 8.61 | 9.72 | 12.11 | | Yutu rover | 1 |
| A11 | 0.73 | 23.49 | 0.19 | 15.8 | 7.5 | 7.81 | 12.01 | 13.45 | 41.86 | 10002, 10010, 10084 | 2, 3 |
| A12 | -2.99 | 336.69 | 0.2 | 15.4 | 3.1 | 9.66 | 10.58 | 13.86 | 45.62 | 12001, 12003, 12023, 12030, 12032, 12033, 12034, 12037, 12041, 12042, 12044, 12057, 12070 | 3, 4 |
| A14 LM-Cone | -3.66 | 342.62 | 0.26 | 10.4 | 1.67 | 9.29 | 11.12 | 17.57 | 47.94 | 14003, 14148, 14149, 14156, 14049, 14163, 14240, 14259, 14421 | 3, 5 |
| A15 LM | 26.14 | 3.67 | 0.22 | 15 | 1.9 | 10.65 | 10.25 | 14.45 | | 15021, 15013 | 6 |
| A15S1-9 | 26.06 | 3.66 | 0.23 | 15.4 | 1.51 | 11.19 | 10.05 | 13.96 | | 15071, 15101, 15201, 15211, 15221, 15231, 15471, 15241, 15261, 15271, 15291, 15301, 15411, 15031, 15041, 15501, 15511, 15531, 15601 | 6, 7 |
| A16S1-9 | -9.03 | 15.49 | 0.38 | 5.5 | 0.61 | 6.04 | 15.51 | 26.67 | 45.07 | 61141, 61161, 61241, 61281, 61501, 62241, 62281, 64421, 64501, 65501, 65701, 65901, 66041, 66081, 68121, 68501, 68821, 68841, 69921, 69941, 69961 | 8 |
| A16S11 | -8.81 | 15.51 | 0.46 | 4.2 | 0.4 | 4.3 | 16.5 | 28.9 | 45.1 | 61141, 61161, 61241, 61281, 61501 | 8 |
| A16S13 | -8.83 | 15.52 | 0.42 | 4.8 | 0.5 | 5.4 | 15.8 | 27.6 | 45.1 | 63321, 63341, 63501 | 8 |
| A17LM | 20.19 | 30.74 | 0.21 | 16.6 | 8.5 | 9.8 | 11.04 | 12.07 | 40.73 | 70019, 70161, 70181, 70011 | 9, 10 |
| A17S1 | 20.16 | 30.75 | 0.2 | 17.8 | 9.6 | 9.62 | 10.75 | 10.87 | 39.93 | 71501, 71041, 71061, 71131, 71151 | 9, 11 |
| A17S3 | 20.17 | 30.53 | 0.3 | 8.7 | 1.8 | 10.25 | 12.89 | 20.29 | 44.94 | 73221, 73241, 73261, 73281 | 10 |
| A17S5 | 20.19 | 30.69 | 0.21 | 17.7 | 9.9 | 9.51 | 10.85 | 10.97 | 39.86 | 75061, 75081 | 9 |
| A17S6-7 | 20.29 | 30.78 | 0.28 | 11.15 | 3.65 | 10.54 | 12.05 | 17.67 | 43.3 | 76241, 76261, 76281, 76321, 76501, 77531 | 9 |
| A17S8 | 20.28 | 30.85 | 0.23 | 12.3 | 4.3 | 9.91 | 11.77 | 15.73 | 42.67 | 78501 | 9 |
| A17LRV7-8 | 20.21 | 30.65 | 0.22 | 15.9 | 6.7 | 10.06 | 11 | 13.1 | 41.85 | 75111, 75121 | 11 |
| A17LRV12 | 20.20 | 30.78 | 0.21 | 17.4 | 10 | 9.36 | 10.7 | 11.15 | 39.9 | 70311, 70321 | 11 |
| Luna16 | -0.71 | 56.37 | 0.18 | 16.7 | 3.3 | 8.8 | 12.5 | 15.3 | 41.7 | | 3 |

| | | | | | | | | | | |
|---|---|---|---|---|---|---|---|---|---|---|
| Luna20 | 3.54 | 56.44 | 0.3 | 7.5 | 0.5 | 9.8 | 15.1 | 22.3 | 45.1 | 3 |
| Luna24 | 12.75 | 62.04 | 0.18 | 19.6 | 1 | 9.4 | 12.3 | 12.5 | 43.9 | 3 |

References: 1. Wu et al. 2018; 2. Rhodes & Blanchard 1981; 3. Heiken et al. 1991; 4. Frondel et al. 1971; 5. Rose et al. 1972; 6. Korotev 1987); 7. Cuttitta (1973; 8. Korotev 1981; 9. Rhodes 1974; 10. Rose 1974; 11. Korotev & Kremser 1992; 12. Korotev et al. 2003.

2.2 Method

The statistical regression model was used to predict the contents of all the six elements. The elemental contents from soil sample stations were plotted against the reflectance values of GF-4, and five univariate regression models, including linear, power, exponential, logarithmic, and polynomial models, were applied to fit the plots with the least square method. Note that obvious outliers would decrease the prediction accuracy, and thus were omitted. The best model was determined according to the squared correlation coefficient ($R^2$). The best-fit lines provided the calibration from the GF-4 reflectance values to absolute major elements contents, and were used to produce the abundance maps of these major elements. The standard deviation (STD) of the elements is defined as (Lucey et al. 2000):

$$\left\{\left[\sum(Element_{predicted} - Element_{actual})^2\right]/(N-1)\right\}^{0.5} \quad (3)$$

where $N$ is the number of stations.

3. Results

3.1 Maps of the major elements

Fig. 2 shows the relationships between the GF-4 reflectance and the abundances of the major elements in the lunar samples established in this study. The GF-4 reflectance decreases with the increasing abundances of FeO, $TiO_2$, and MgO, but increases along with the increase of the $Al_2O_3$, CaO, and $SiO_2$ abundances. The best-fit models are power for FeO, $TiO_2$, and $SiO_2$, exponential for MgO, and linear for CaO and $Al_2O_3$. Except for $SiO_2$, all the $R^2$ are ≥0.87, and even greater than 0.92 for FeO, CaO, and $Al_2O_3$, indicating that it fits well. Fig. 3 shows the plots of predicted versus measured abundances of FeO and $Al_2O_3$ at the sample stations. Both the slope and the $R^2$ are >0.9, indicating the model used in this study achieves a relatively good performance in predicting the elements contents. The STDs for FeO and $Al_2O_3$ are 1.58 and 1.46 respectively, comparable with that in Lucey et al. (2000) and Wu (2012).

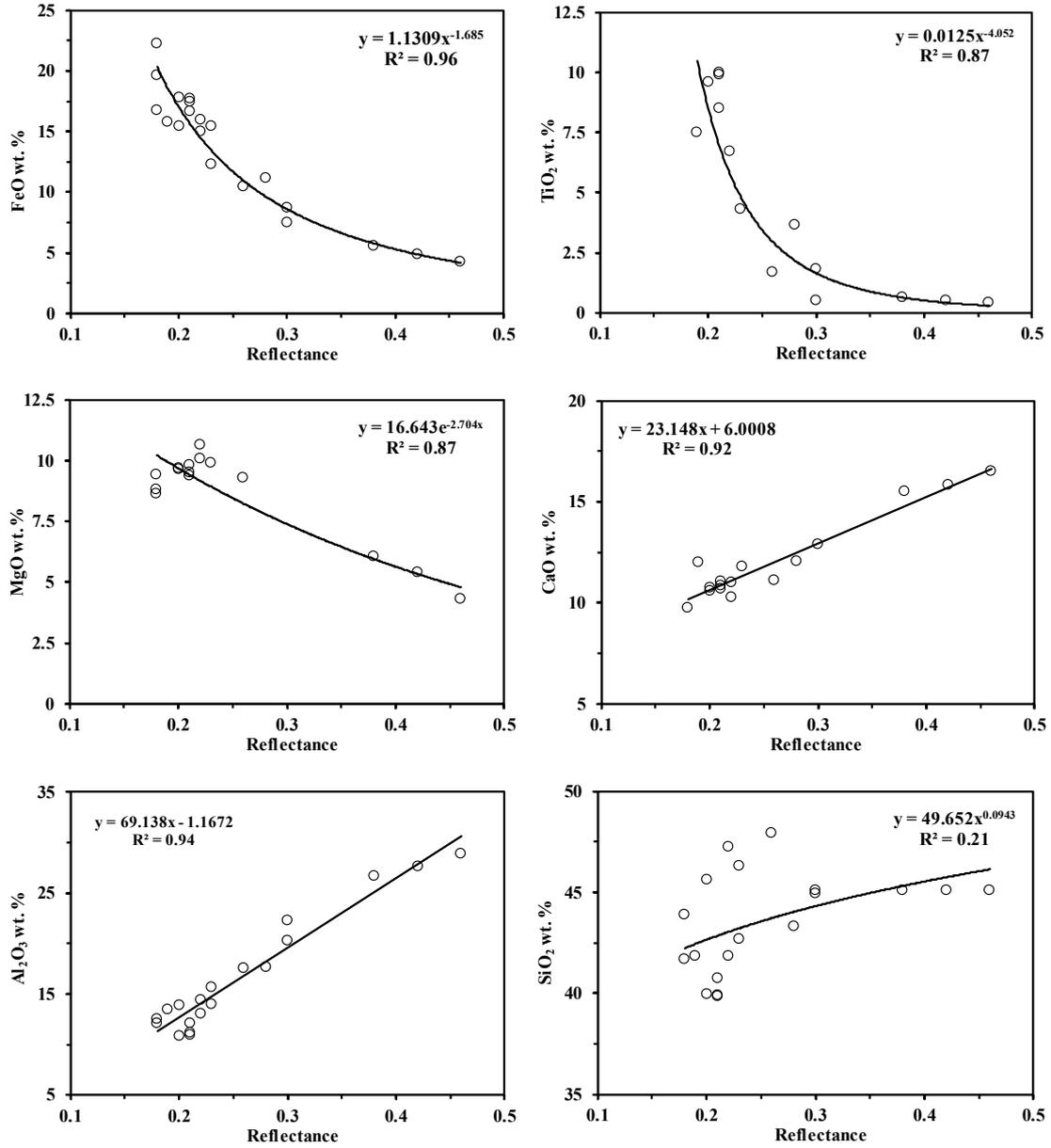

Fig. 2. The relationships between the GF-4 band 5 reflectance and the abundances of the six major elements of the samples in Table 1.

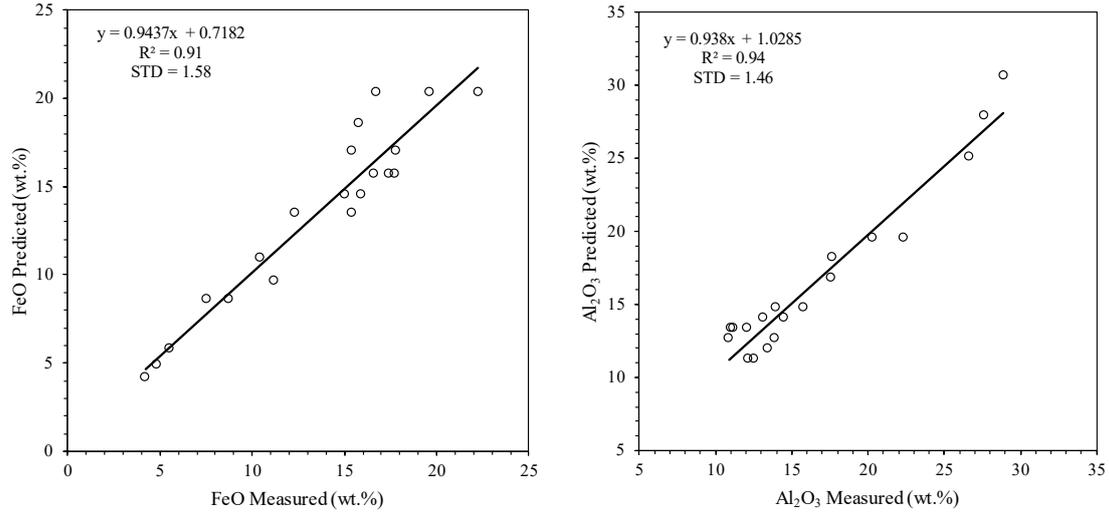

Fig. 3. Plots of predicted vs measured abundances of FeO and Al$_2$O$_3$.

Fig. 4 shows the maps of the elemental contents on the lunar nearside produced using the models established above. Compared with previous maps produced by Clementine (Lucey et al. 1995, 2000; Blewett et al. 1997; Gillis et al. 2004, 2006), M$^3$ (Bhatt et al. 2019), or Chang'E-1 IIM (Wu 2012; Xia et al. 2019), it is obvious that the maps derived in this study are seamless and homogenous, without hue variations or gaps (Fig. 5). The abundances of the major elements in this study are comparable with previous results. The abundance of FeO is ~5.5-20.1 wt.% in this study, and it is ~2.1-16.4 wt.% in Wu (2012), ~0-20 wt.% in Lucey et al. (2000) and Wu et al. (2012), ~2.5-22.5 wt.% in Lemelin et al. (2016). The abundances of TiO$_2$ in this study is ~0.6-12.6 wt. %, and it is ~0-8 wt.% in Wu (2012), ~2-10 wt.% in Sato et al. (2017). Note that the maps in this study only covered the lunar nearside, there may exist some differences, e.g., the lower limit of the abundance of FeO is a little higher than other results.

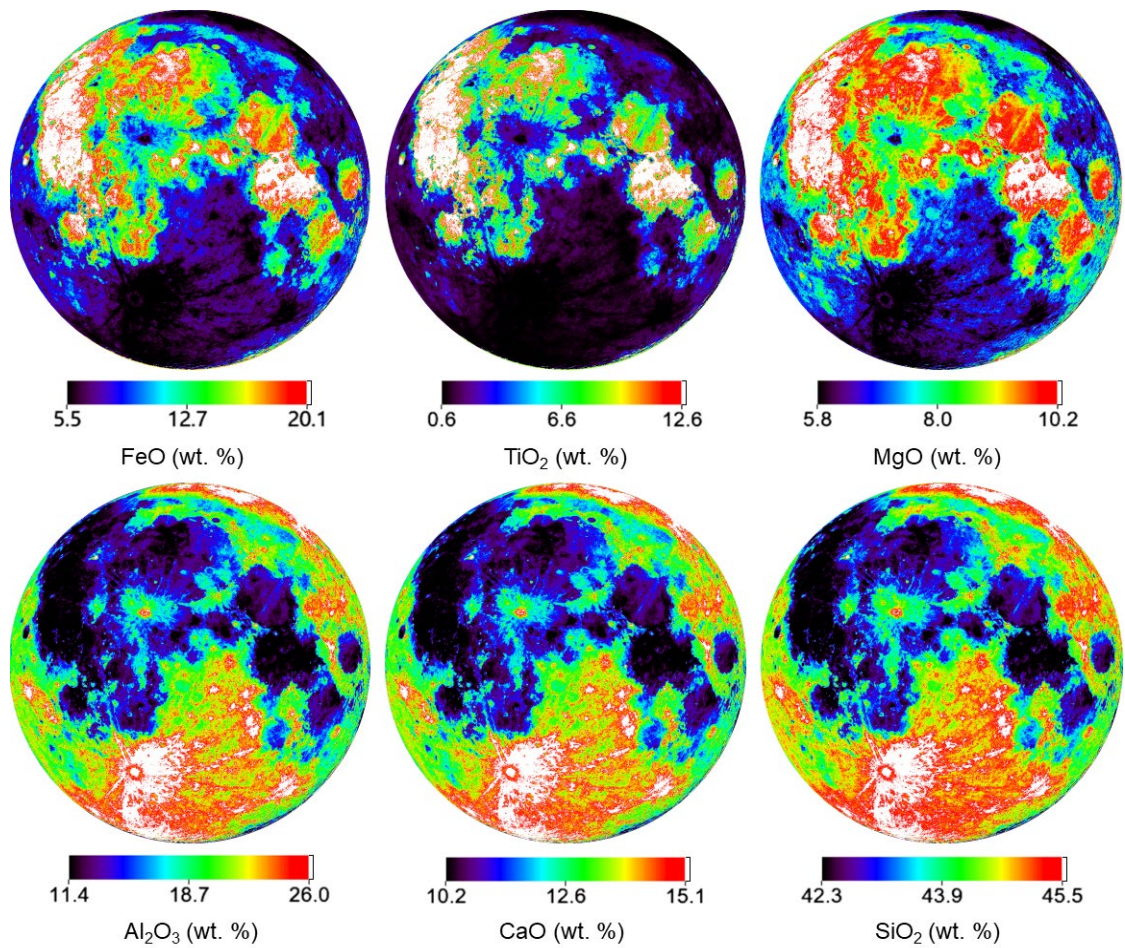

Fig. 4. Maps of the abundances of the major elements on the lunar nearside derived from GF-4 data.

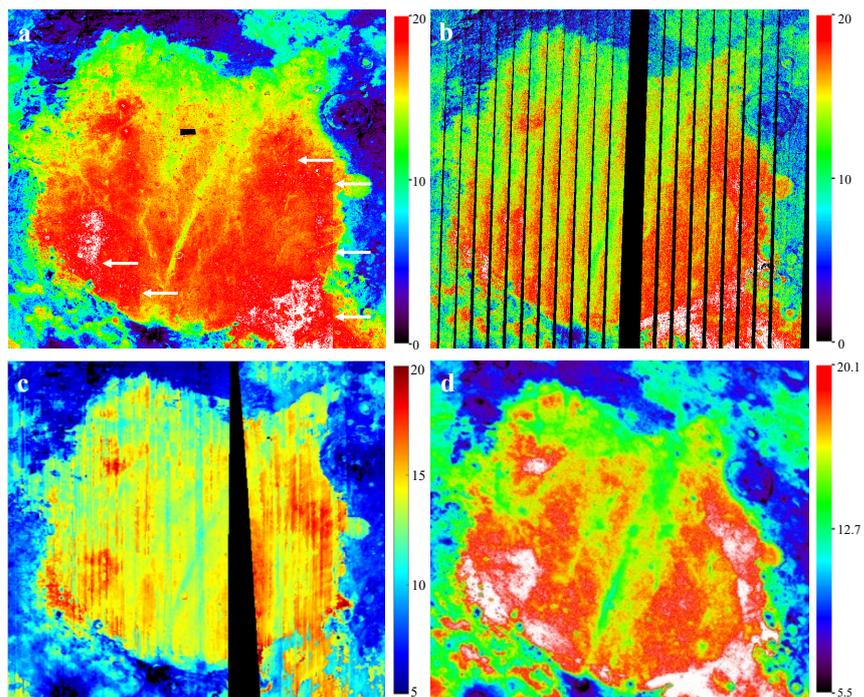

Fig. 5. Comparison of FeO abundances in different data sets in Mare Serenitatis region. (a) FeO abundance map derived from Clementine (Lucey et al. 2000). (b) FeO abundance map derived from Chang'E-1 IIM (Wu 2012). (c) FeO abundance map derived from M$^3$ (modified from Bhatt et al. (2019)). (d) FeO abundance map derived from GF-4. White arrows in (a) show the hue variations around the mosaic borders, which are very prominent in (b) and (c). Black cubes indicate the gaps due to lack of data.

The spectrally unique and unsampled Eratosthenian basalts distributed in Mare Imbrium (including the Chang'E-3 landing zone), Oceanus Procellarum, and Mare Tranquillitatis are prominent in the maps of elemental contents (Fig. 4). These basalts have the highest FeO, TiO$_2$, and MgO abundances (white color in Fig. 4), different from other mare basalts. The high resolution maps of these major elements produced in this study clearly show these differences, and thus are beneficial for geologic studies, e.g., division of different geologic units.

Maria, such as Mare Imbrium, Oceanus Procellarum, Mare Serenitatis, Mare Tranquillitatis, are generally rich in FeO, TiO$_2$, and MgO. However, Mare Frigoris is an exception. Compared with other maria, the average TiO$_2$ content in Mare Frigoris (~2.91 wt.%) is much lower, and the average Al$_2$O$_3$ content (~17.24 wt.%) is higher, indicating that most of the basalts in Mare Frigoris are high-Al basalts (Kramer et al. 2015).

The abundance maps of the major elements, especially Al$_2$O$_3$, CaO, and SiO$_2$, clearly show the ejecta distribution of Tycho crater (Fig. 4). The bottom and eastern part of Tycho crater are rich in Al$_2$O$_3$, CaO, and SiO$_2$, while the contents of these elements on the crater wall and the western part of the crater are relatively lower. The asymmetric distributions of these major elements around Tycho crater may suggest that Tycho crater was formed by an oblique impact from the southwest direction, consistent with the studies on the distributions of impact melt and secondary craters of Tycho (Hirata et al. 2004; Krüger et al. 2016). The abundances of Al$_2$O$_3$, CaO, and SiO$_2$ on the relatively fresh ejecta of Tycho crater are higher than the abundances on the surrounding highland areas, which may be due to space weathering effects.

3.2 Statistical analysis

Fig. 6 shows the abundance distributions of all the six major elements. Except for $TiO_2$, which shows unimodal continuum distribution, all other elements exhibit bimodal distributions, corresponding to the maria and highlands respectively, consistent with previous studies (Lucey et al. 1998; Giguere et al. 2000; Gillis et al. 2004; Wu 2012; Wu et al. 2012). The lower modal Fe abundance of ~6.65 FeO wt.%, the higher modal Fe abundance of ~16.31 FeO wt.%, which corresponds to the highland areas and mare areas respectively, are little higher than the abundance of ~5.57 wt.% given by Wu (2012), or ~5.7 wt.% by Gillis et al. (2004), consistent with the fact that the maps in this study only covered the lunar nearside, and the abundances of FeO on the lunar farside (mostly highland areas) are much lower than on the nearside. Besides, the average abundances of these major elements in maria, highlands and the whole lunar nearside were estimated and compared (Fig. 7). The abundances of FeO, $TiO_2$, and MgO in the maria are higher than those in the highlands, while the abundances of $Al_2O_3$, CaO, and $SiO_2$ are opposite. Fig. 8 shows the proportion of the abundances of the major elements in maria, highlands, and the whole lunar nearside. As can be seen, both in maria and highlands, the $SiO_2$ has the highest proportion (> 40%) among these major elements, while the $TiO_2$ has the lowest proportion (< 10%), e.g., $TiO_2$ only makes up 1.29% of elements in the highlands.

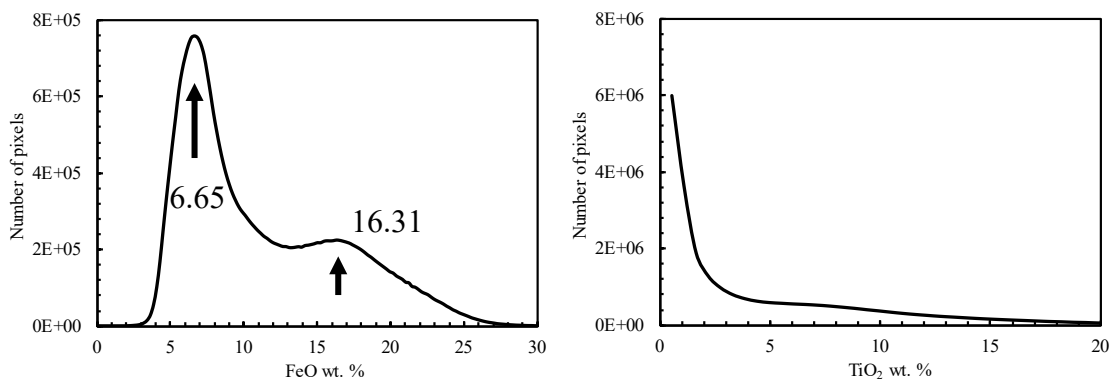

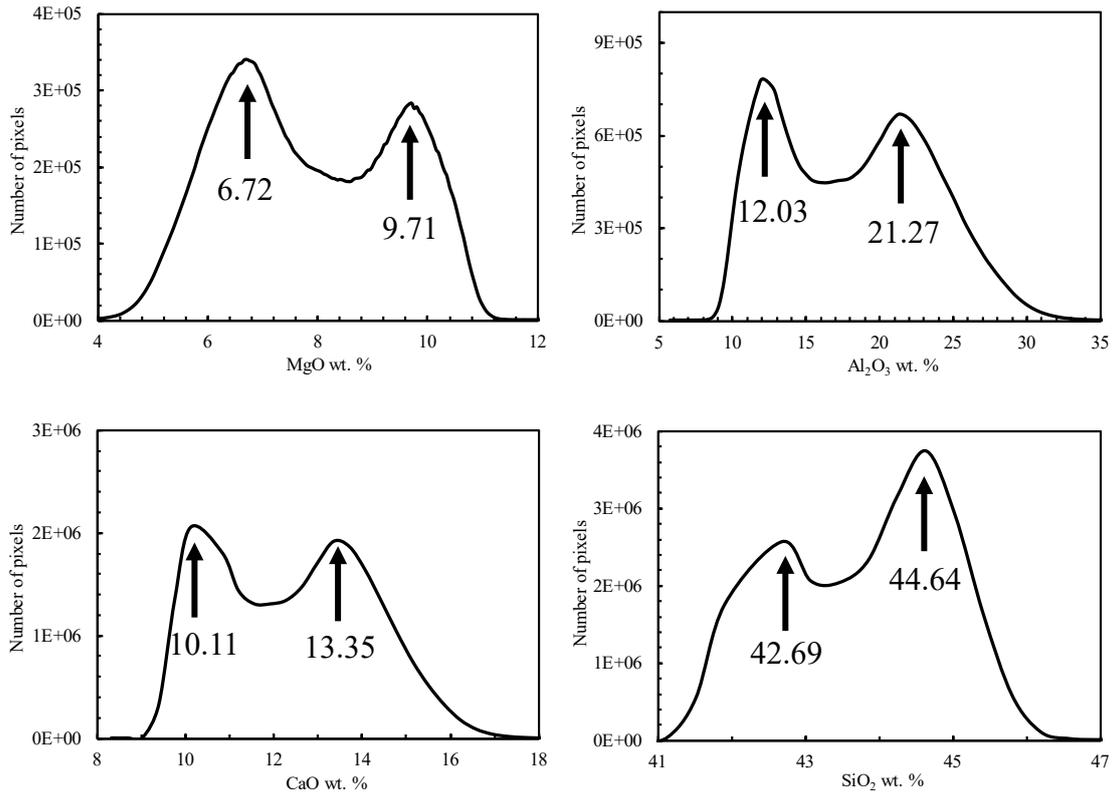

Fig. 6. Histograms of elemental abundances for the lunar nearside derived from GF-4 data.

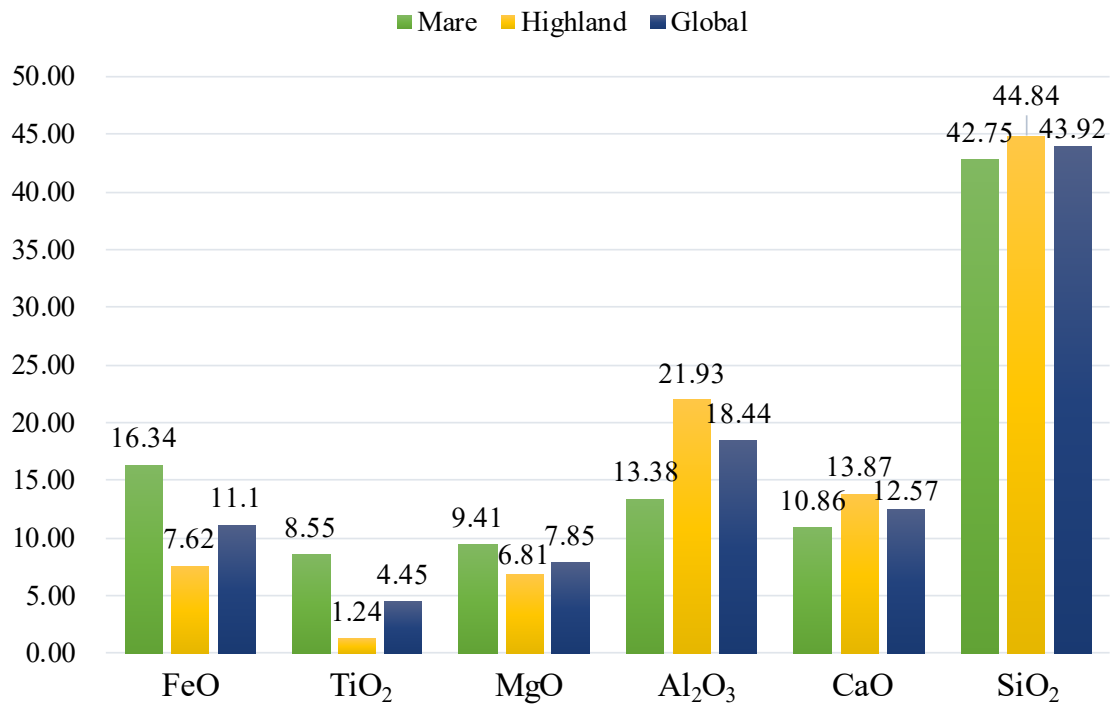

Fig. 7. Average abundances of the six major elements derived from GF-4 data.

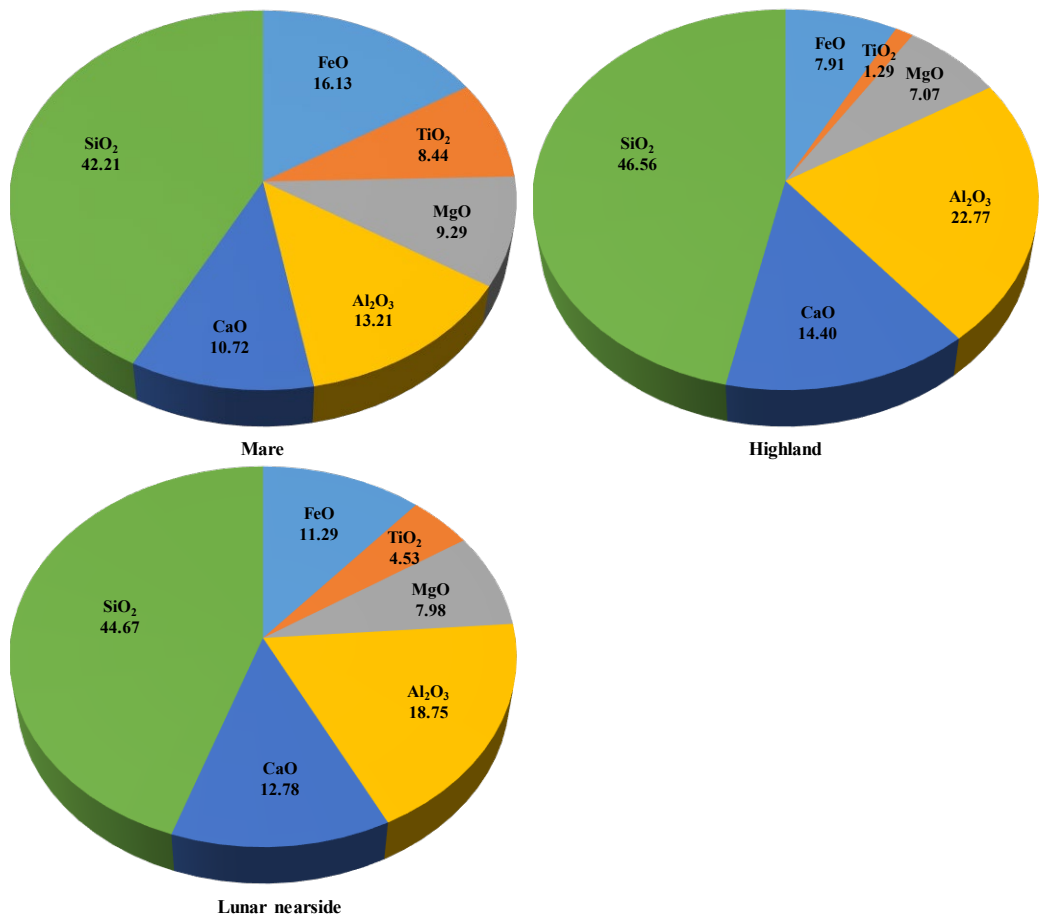

Fig. 8. Pie chart showing the proportion of the abundances of the major elements in maria, highlands, and the lunar nearside.

4. Conclusions

China's high-resolution geostationary satellite, Gaofen-4, imaged the whole lunar disk with spatial resolution of ~500 m in single-exposure. Using these data this study produced the seamless and homogeneous distribution maps of the abundances of the major elements (Fe, Ti, Mg, Al, Ca, Si) on the lunar nearside. Compared with previous maps, the maps derived in this study show no hue variations and gaps, and thus are more beneficial for geologic studies, e.g., division of different geologic units. With these products the average contents and proportions of the major elements for mare and highland areas were estimated and compared. The results showed that both in maria and highlands, $SiO_2$ has the highest proportion among these major oxides, while $TiO_2$ has the lowest. The asymmetric distributions of $Al_2O_3$, CaO, and $SiO_2$ around Tycho crater may suggest that Tycho crater was formed by an oblique impact from the southwest direction. Some related geological researches will be carried out in the future based on the products produced in this study.

Acknowledgments:

We would like to thank China National Space Administration (CNSA) for organizing the observations. This work was supported by the pre-research project on Civil Aerospace Technologies of China National Space Administration (Grant No. D020203) and Minor Planet Foundation of Purple Mountain Observatory.